\documentclass[aps,prd,superscriptaddress,twocolumn]{revtex4}
\usepackage{epsfig}
\usepackage{graphicx}
\usepackage{amsmath,amssymb,amsfonts,latexsym}
\usepackage{graphicx}

\usepackage{epsfig,amssymb,amsfonts,verbatim}

\usepackage{amsmath}
\usepackage{latexsym}
\usepackage{amsfonts}
\usepackage{amssymb}
\usepackage{color}

\def\bfl{\begin{flushleft}}
\def\efl{\end{flushleft}}
\def\bfr{\begin{flushright}}
\def\efr{\end{flushright}}
\def\bc{\begin{center}}
\def\ec{\end{center}}

\def\ba{\begin{eqnarray}}
\def\ea{\end{eqnarray}}
\def\baa#1{\begin{array}{#1}}
\def\eaa{\end{array}}
\def\bw{\begin{widetext}}
\def\ew{\end{widetext}}

\def\text#1{\mbox{#1}}

\begin{document}


\title{Non-square-well potential profile and suppression of blinking in compositionally graded Cd$_{1-x}$Zn$_x$Se/Cd$_x$Zn$_{1-x}$Se nanocrystals}

\author{Andrew Das Arulsamy}
\email{sadwerdna@gmail.com}
\affiliation{School of Physics, The
University of Sydney, Sydney, New South Wales 2006, Australia}
\affiliation{Jo$\check{z}$ef Stefan Institute, Jamova cesta 39, SI-1000 Ljubljana, Slovenia, EU}

\author{Uros Cvelbar}
\affiliation{Jo$\check{z}$ef Stefan Institute, Jamova cesta 39, SI-1000 Ljubljana, Slovenia, EU}
\author{Miran Mozetic}
\affiliation{Jo$\check{z}$ef Stefan Institute, Jamova cesta 39, SI-1000 Ljubljana, Slovenia, EU}
\author{Kostya (Ken) Ostrikov}

\affiliation{CSIRO Materials Science and Engineering, P.O. Box 218, Lindfield NSW 2070, Australia}
\affiliation{School of Physics, The University of Sydney, Sydney, New South Wales 2006, Australia}

\date{\today}

\begin{abstract}
Random blinking is a major problem on the way to successful applications of semiconducting nanocrystals in optoelectronics and photonics, which until recently had no practical solution nor theoretical interpretation. An experimental breakthrough has recently been made by fabricating non-blinking Cd$_{1-x}$Zn$_x$Se/ZnSe graded nanocrystals [Nature \textbf{459}, 686 (2009)]. Here, we report (1) an unequivocal and detailed theoretical investigation to understand the properties (e.g., profile) of the potential-well and the distribution of Zn content with respect to the nanocrystal radius and (2) develop a strategy to find the relationship between the photoluminescence (PL) energy peaks and the potential-well due to Zn distribution in nanocrystals. It is demonstrated that the non-square-well potential can be varied in such a way that one can indeed control the PL intensity and the energy-level difference (PL energy peaks) accurately. This implies that one can either suppress the blinking altogether, or alternatively, manipulate the PL energy peaks and intensities systematically to achieve a controlled non-random intermittent luminescence. The approach developed here is based on the ionization energy approximation and as such is generic and can be applied to any non-free-electron nanocrystals.
\end{abstract}


\maketitle

\subsection*{1.~~Introduction}

Quantum dots (QDs) in all fields of applications have one intrinsic problem$-$ they blink randomly. Blinking is defined by intensity fluctuations in nanocrystals (NCs) or QDs, which is due to intermittent photoluminescence (PL) under continuous photoexcitation~\cite{wangnat}. This blinking effect is due to the fluctuations of PL from an ensemble of individual NCs. In other words, the excited electrons radiatively recombine with holes at different rates in individual NCs. Such discontinuous PL causes some of the QDs to be switched-on (emission) while others remain switched-off (due to trapped electrons or holes and also due to non-radiative Auger process)~\cite{wangnat,kli,coon}. The trapped electrons have relatively large recombination lifetimes, which vary from one NC to another in an ensemble of NCs. The blinking effect was first observed and reported by Nirmal et al.~\cite{nirmal}, and has been discussed extensively since then~\cite{gom,wang,raj,dunn}. Interested readers are referred to Ref.~\cite{dunn} for a thorough review on the blinking effect.

These discontinuous emissions are indeed undesirable for solar cells, nano-electronics and biological applications. Hence, one needs to fine-tune the electronic properties of the QDs so as to avoid blinking in the QDs~\cite{wangnat}. These random and intermittent emissions in a single NC that give rise to the blinking effect, have been suppressed recently by custom-designing a compositionally graded CdZnSe/ZnSe core/shell structure~\cite{wangnat}. These graded structures have given rise to a series of smaller energy levels for the excited electrons to recombine at a faster rates within the NCs, which in turn resulted in quasi-continuous emission. Therefore, further theoretical work need to be carried out to understand (I) why such compositionally graded structures have increased the recombination rates and (II) how one can further improve the grading and the electron confinement in other material systems and core/shell structures to eliminate the blinking effect. 

Apart from non-blinking effect, the points (I) and (II) stated above are also crucial to the advancement of nanotechnologies in different fields such as renewable energy, nano-electronics and biomedicine. For example, one of the straightforward applications of QDs is in photovoltaic solar cells. In the latter application, one does not need to control the size and spatial distribution of QDs accurately since one can achieve an energy gap distribution arising from the size non-uniformity of QDs. In fact, such non-uniformity enables effective photon energy absorption (between 0.5 to 3.5 eV) from the sunlight~\cite{green,cho,ser}. However, the carrier multiplication is not effectively enhanced even in the presence of high energy photons~\cite{na1,na2}. Meanwhile, studies of the electron confinement and PL in Si nanostructures~\cite{jul,sy,ciu} reveal that small Si-based QDs absorb higher-energy photons, whereas larger QDs absorb photons with lower energy. Clearly, fluctuating recombination rates in these QDs are undesirable for a continuous photovoltaic power source. On the other hand, QD research fields related to biological applications are also emerging and expanding exponentially~\cite{biju1,cot}, where QDs have been successfully used as bio-probes~\cite{fab} and also in biological imaging~\cite{par,jyo}. Such biological applications give rise to the need to study the effect of QD chemical properties and their nanoscale size on health and environment, which have been pointed out and emphasized in the recent years~\cite{biju,par,ho,ban,bar}. Many of the physical phenomena that enable such applications are essentially based on the points (I) and (II) above. 

Here, we will analyze and understand the variation of the energy-level spacing responsible for the non-blinking effect in the Cd$_{1-x}$Zn$_x$Se/ZnSe NCs with respect to Zn content, $x$ (elemental composition). Our objectives to achieve points (I) and (II) mentioned above are (a) to re-interpret the experimental results reported in Ref.~\cite{wangnat}, and (b) to make experimentally viable predictions for similar effects in other quantum dot and materials systems. The analysis presented here will also enable one to predict the PL spectra for the core/shell structure of ZnSe/Cd$_{1-x}$Zn$_x$Se NCs, an inverse NC structure as compared to CdZnSe/ZnSe~\cite{wangnat}. In (a) we will attempt to interpret the experimentally measured PL intensities and the energy peaks as a function of Zn content ($x$). We will also develop generic guidelines on the suppression of blinking effects in multi-element nanocrystals made of non-free-electron materials. 

The paper is organized as follows. The details of our model with respect to the ionization energy theory (IET) are given in Section 2. In Section 3, we present the detailed analysis on the Cd$_{1-x}$Zn$_x$Se/ZnSe nanocrystals. In Section 4, we discuss and make predictions on the ZnSe/Cd$_{1-x}$Zn$_x$Se, an inverse core/shell structure of Cd$_{1-x}$Zn$_x$Se/ZnSe nanocrystal. The analysis in Section 5 focuses on the effect of grading-depth due to inhomogeneous Zn concentration in these NCs. In addition, we also evaluate the possibility of achieving the sufficient grading-depth for other well-known systems, namely, CdSe/ZnS and InGaAs/GaAs. The paper ends with a brief concluding section, where the main results are summarized.      

\subsection*{2.~~Model}

Our model for describing the potential wells in elementally graded NCs is based on the IET, which relates the atomic ionization energy of the constituent atoms to the energy level difference of NCs. The Schr$\ddot{\rm o}$dinger equation for the IET is given by~\cite{a2}

\begin {eqnarray}
\hat{H}\varphi = (E_0 \pm \xi)\varphi. \label{eq:1}
\end {eqnarray}

The microscopic and mathematical details of the Hamilton operator, $\hat{H}$ and the IET can be found in Ref.~\cite{a3}. The exact eigenvalue is given by $E_0 \pm \xi$, where $E_0$ is the total energy of the system at zero temperature ($T = 0$ K) and $\xi$ is the energy-level difference (also called the ionization energy) in a given NC, which is equivalent to the energy peak positions in the PL spectra. By identifying $\xi$ as the real energy level difference in NCs and $E_I$ as the average atomic energy level differences (averaged from all the constituent atoms in NCs), one can write the eigenvalue in Eq.~(\ref{eq:1}) as 

\begin {eqnarray}
E_0 \pm \xi = E_0 \pm \beta\sum_i^z\frac{E_{Ii}}{z} \propto E_0 \pm \sum_i^z \frac{E_{Ii}}{z}, \label{eq:2}
\end {eqnarray}

where the subscript $i$ counts the first, second, ..., $z$ ionization energy of each constituent atom for a given material. Here, $\sum_i^z E_{Ii}/z$ gives the changes to the average ionization energy of a given NC system. In Eq.~(\ref{eq:2}), the coefficient $\beta$ is defined as

\begin {eqnarray}
\beta = 1 \pm \frac{\langle V^{\rm many}_{\rm body}\rangle}{\sum_i^z E_{Ii}/z}, \label{eq:3}
\end {eqnarray}
 
where $\langle V^{\rm many}_{\rm body}\rangle$ is the many-body potential. This potential may increase the real energy level difference in a given NC or solid, compared to the atomic ionization energy ($\xi > E_I$), or may decrease the real energy level difference, so that $\xi < E_I$. Therefore, the label ``+" in ``$\pm$" from Eq.~(\ref{eq:3}) implies $\xi > E_I$, whereas ``$-$" implies $\xi < E_I$. For example, for electrons, Eq.~(\ref{eq:2}) can also be written as (using Eq.~(\ref{eq:3}))

\begin {eqnarray}
&E_0 + \xi & = E_0 + \sum_i^z \frac{E_{Ii}}{z} \pm \langle V^{\rm many}_{\rm body}\rangle. \label{eq:4}
\end {eqnarray}

It is clear that $\xi \geq E_I$ is valid for $\beta \geq 1$, while $\xi < E_I$ is valid for $0 < \beta < 1$. In the subsequent analysis, Eqs.~(\ref{eq:2}) $-$ (\ref{eq:4}) will be used to interpret the PL spectra for Cd$_{1-x}$Zn$_x$Se/ZnSe NCs. For example, Eq.~(\ref{eq:4}) indicates that $E_0$ and $\langle V^{\rm many}_{\rm body}\rangle$ are material-specific constants for one given composition, and any changes to the composition by varying the Zn content, $x$ can be directly related to $E_{I}$. Therefore, any changes to the elemental composition can be related to the $E_{I}$ and, subsequently, to the intensities and energy peak positions in the PL spectra.      

\subsection*{3.~~Analysis I: Cd$_{1-x}$Zn$_x$Se/ZnSe}    

\begin{figure}[hbtp!]
\begin{center}
\scalebox{0.37}{\includegraphics{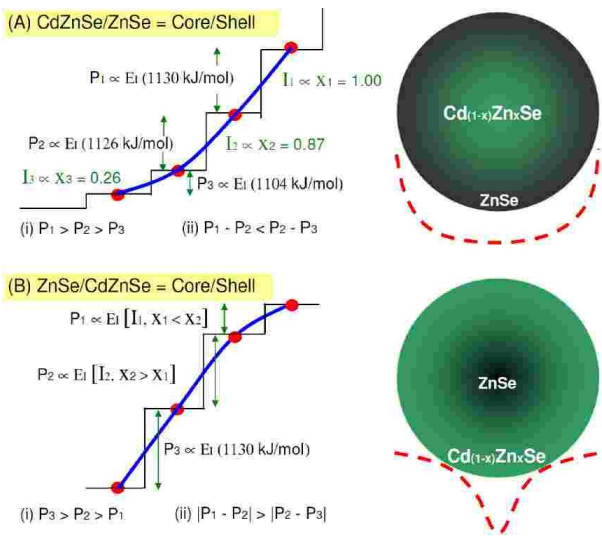}}
\caption{Panel (A) shows the one-half potential (solid line) for the core/shell structure, CdZnSe/ZnSe. The step-like feature represents the actual discreet energy levels in the nanocrystals. Panel (B) shows the expected one-half potential well if we were to have ZnSe/CdZnSe as core/shell, i.e., the inverse structure of panel (A). The full potential profiles are drawn with dashed lines for both (A) and (B). Both panels (A) and (B) show the sketches of the confining potentials for the respective NCs; there is only one confining potential for each graded or ungraded NC. Therefore, the three different peaks correspond to the three different values of the energy-level spacing within the same confining potential rather than three different confining potentials. These schematic diagrams are not to scale.}
\label{fig:1}
\end{center}
\end{figure}

Figure~\ref{fig:1}A indicates the expected profile of the graded potential (almost a parabolic well) as measured in Ref.~\cite{wangnat} for a Cd$_{1-x}$Zn$_x$Se/ZnSe core/shell NC structure. The data from PL spectra reported by Wang et al.~\cite{wangnat} can be written in such a way that the PL spectra satisfy two conditions, (i) $P_1$ $>$ $P_2$ $>$ $P_3$ and (ii) $P_1$ $-$ $P_2$ $<$ $P_2$ $-$ $P_3$, where $P$ denotes the measured PL energy peak positions. We also label the PL peak intensities as $I_1$, $I_2$ and $I_3$, respectively. We now interpret the intensity, $I_1$ for the PL peak $P_1$, which corresponds to the Zn content $x_1$; in other words, $x_1 \propto I_1$, $x_2 \propto I_2$ and $x_3 \propto I_3$. Hence, our interpretation reads, $x_1 = 1.00 \propto I_1$, $x_2 = 0.87 \propto I_2$ and $x_3 = 0.26 \propto I_3$, see Fig.~\ref{fig:1}A for details. The values 1.00, 0.87 and 0.26 are the normalized PL intensities imported from Ref.~\cite{wangnat}.

Conditions (i) and (ii), after invoking the IET give rise to the potential-well profile as shown in Fig.~\ref{fig:1}A. Note here that we only show the one-half of the potential well for convenience. The step-like feature in Fig.~\ref{fig:1}A is entirely due to the changes in the ionization energy (energy-level spacing) as a result of changing $x$ with radius, $r$ ($x(r)$) of the NC from the center ($r = r_0$ = 0) or from the outer surface ($r$ = $r_{\rm NC}$). This step-like (discreet energy levels) feature can be represented with a continuous solid line given in Fig.~\ref{fig:1}A that represents the one-half of the potential well (the full potential profiles are denoted with dashed lines). 

\begin{table}[ht]
\caption{Averaged atomic ionization energies ($E_I$) for individual ions. These ions are arranged with increasing atomic number $Z$. Note here that the elements, S and Se are anions, and therefore in the IET calculations one only requires to know their first ionization energies. The unit kJ/mol is adopted for numerical convenience.} 
\begin{tabular}{l c c c } 
\hline\hline 
\multicolumn{1}{l}{Ion}            &    ~~~Atomic number  & ~~~Valence    & ~~~$E_I$   \\  
\multicolumn{1}{l}{}                &   ~~~$Z$             & ~~~state      & ~~~(kJ/mol)\\  
\hline 

S                                   &  16   	  			  &  1+      & 1000 \\ 
Zn                                  &  30  					    &  2+      & 1320 \\ 
Ga                                  &  31 					    &  3+      & 1840 \\
As                                  &  33	  				    &  3+      & 1827 \\ 
Se                                  &  34  					    &  1+      & 941  \\ 
Cd                                  &  48 					    &  2+      & 1250 \\ 
In                                  &  49 					    &  3+      & 1694 \\ 

\hline  
\end{tabular}
\label{Table:I} 
\end{table}

Recall here that in contrast to Ref.~\cite{wangnat}, we did not assume $P_1 - P_2 = P_2 - P_3$ because this assumption would not be consistent with the results of the PL measurements for all the NCs studied in Ref.~\cite{wangnat}. This experiment, as well as our IET model, suggest that the emission from all the NCs always satisfies $P_1 - P_2 < P_2 - P_3$. Indeed, the difference between the peak positions $P_1 - P_2$ and $P_2 - P_3$ is quite different in our model and the experiment. This difference may be due to the qualitative nature of the IET approximation and values of the ionization energy used. Nonetheless, what is most important here is that both sets of data do satisfy the essential condition $P_1 - P_2 < P_2 - P_3$ and as such, are consistent.

The ionization energy (see Table~\ref{Table:I}) for the ZnSe shell ($E_I^{\rm shell}$) can be approximated (ionization energy approximation~\cite{a2,a3}) as $E_I^{\rm shell}$ = $\frac{1}{2}$Zn + $\frac{1}{2}$Se = $\frac{1}{2}$(1320) + $\frac{1}{2}$(941) = 1130 kJ/mol. On the other hand, the ionization energy for the core, Cd$_{1-x}$Zn$_x$Se is $E_I^{\rm core}$ = $\frac{1}{2}$[($x$)Zn + (1$-x$)Cd] + $\frac{1}{2}$Se. Therefore, for $x_1$ = 1.00, $E_I^{\rm shell}$ = $E_I^{\rm core}$ = 1130 kJ/mol, for $x_2$ = 0.87, $E_I^{\rm core}$ = $\frac{1}{2}$[(0.87)(1320) + (0.13)(1250)] + $\frac{1}{2}$(941) = 1126 kJ/mol. For $x_3$ = 0.26, we obtain, $E_I^{\rm core}$ = 1104 kJ/mol. 


We can now recall the PL intensities and rewrite them as functionals of $x(r)$: $I_1[x_1(r_1)]$, $I_2[x_2(r_2)]$ and $I_3[x_3(r_3)]$, where $r_1 > r_2 > r_3$. As a consequence, smaller intensity implies lower content of Zn or smaller $x$ due to decreasing $r$ (radially moving inward from the shell to the core of the NC). The existence of such functionals have been shown experimentally in Y$_{x}$Gd$_{1-x}$VO$_{4}$:Eu$^{3+}$ by Wu and Yan~\cite{wu}. Furthermore, three-dimensional numerical simulations (for a fixed carrier density: 10$^{22}$ cm$^{-2}$s$^{-1}$) were carried out by Benbakhti et al.~\cite{ben} resulting in the carrier-density dependent (or $x$-dependent in our case) PL intensities.

Consequently, one can explain the origin of the conditions, (i) and (ii) discussed above using the ionization energy approximation. The reason for (i) is due to decreasing Zn content or $x$ (as one moves inwards into the NC) that gives rise to decreasing ionization energies from 1130 to 1126 kJ/mol and then to 1104 kJ/mol. Note here that the ionization energy is the atomic energy level difference (or the energy-level spacing). Condition (ii) is also satisfied: 1130 $-$ 1126 = 4 kJ/mol = 41 meV/atom and 1126 $-$ 1104 = 22 kJ/mol = 228 meV/atom, thus, 41 $<$ 228 meV/atom. From Ref.~\cite{wangnat}, condition (ii) reads $P_1 - P_2$ = 156 meV, $P_2 - P_3$ = 171 meV and therefore, 156 $<$ 171 meV. 

All the ionization energy values prior to averaging were taken from Ref.~\cite{web}. The IET approximation has been shown to be accurate in non-free-electron solids of any dimensions (from zero (QD)- to three-dimensional (bulk) materials) that can be used to understand the properties of strongly correlated matter~\cite{a3}. Furthermore, it is also worth to mention that there are reports on the growth of graded Si$_{1-x}$C$_x$, InSb and GaAs QDs via the plasma-assisted nano-assembly~\cite{k1,k2,k3,k4}. In these studies, it was shown that the grading of elemental composition in the QDs can be controlled systematically. Interestingly, the plasma-assisted growth mechanism have been successfully implemented experimentally to synthesize the Si$_{1-x}$C$_x$ QD arrays~\cite{k6} and iron oxide nanostructures~\cite{k57}. Further details on this experimental and numerical techniques can be found in the Refs.~\cite{k5,k7,k8}. In the following section, we will study and predict the PL properties of the ZnSe/Cd$_{1-x}$Zn$_x$Se structure.

\subsection*{4.~~Analysis II: ZnSe/Cd$_{1-x}$Zn$_x$Se}    

In previous sections, we have explained the essential experimental PL results presented in Ref.~\cite{wangnat}, namely the changes of PL intensities, conditions (i) and (ii). The next step is that we need to use this information to further elaborate the approach to suppress blinking in NCs of different elemental compositions. These predictions are particularly important for the development of new and improved non-blinking nanocrystalline materials. We also propose the possibility to vary the potential well of any non-free-electron NCs and/or QDs at will by simply changing the elemental grading, $x(r)$. The changes to this potential well can be estimated accurately by measuring the intensities, the PL energy peak positions [condition (i)] and the energy difference between the PL energy peak positions [condition (ii)]. 

For example, we expect the potential well for the core/shell ZnSe/Cd$_{1-x}$Zn$_x$Se structure (Fig.~\ref{fig:1}B) to be different compared to the core/shell structure for the Cd$_{1-x}$Zn$_x$Se/ZnSe NC discussed earlier (Fig.~\ref{fig:1}A). The reason is that $E_I^{\rm shell} < E_I^{\rm core}$, and this inequality implies that the ionization energy increases with the depth of the NC, or as one moves inwards (decreasing $r$) into the NC. In the case discussed earlier with the Cd$_{1-x}$Zn$_x$Se/ZnSe NC, we had $E_I^{\rm shell} > E_I^{\rm core}$. The latter inequality implies decreasing ionization energy due to decreasing Zn content as one approaches the center of the NC. Recall here that $E_I^{\rm Zn} > E_I^{\rm Cd}$. This last inequality together with the dependence $x(r)$ define the profile of the potential well in these NCs. By controlling the $x(r)$ accurately via diffusion or by any other means one should be able to control both the potential well profile and hence the strength of the electron confinement in NCs. In the following section, $x(r)$ will be revisited with further analysis.

\subsection*{5.~~Further analysis and predictions for other nanocrystalline systems}    

Let us now use Eqs.~(\ref{eq:2})$-$(\ref{eq:4}) to demonstrate that the $\beta$ [Eq.~(\ref{eq:3})] for NCs is bounded in $0 < \beta < 1$, in which $\xi < E_I$. It is clear from the above discussion that the real energy level differences ($\xi$) in the NC, Cd$_{1-x}$Zn$_x$Se/ZnSe satisfy the inequality, $\xi$(156 and 171 meV/NC) $<$ $E_I$(41 and 228 meV/atom). Therefore, Eq.~(\ref{eq:4}) can be rewritten as   

\begin {eqnarray}
E_0 + \xi = E_0 + \sum_i^z \frac{E_{Ii}}{z} - \langle V^{\rm many}_{\rm body}\rangle, \label{eq:5}
\end {eqnarray}

where $0 < \beta < 1$. From Eq.~(\ref{eq:2}) and the definition of Eq.~(\ref{eq:3}), we obtain $\beta_{12}$ = (156/41) atom/NC and $\beta_{23}$ = (171/228) atom/NC. Since the number of atoms in a $\sim$5 nm diameter NC in Ref.~\cite{wangnat} is definitely much larger than 10 atoms and therefore $\beta_{12}$ and $\beta_{23}$ are indeed bounded in $0 < \beta < 1$. This means that the potential well is sensitive to the small changes to the elemental composition $x$, or the Zn content. Here, $\beta = 0$ gives rise to free-electron system, while $\beta = 1$ implies that one can make accurate quantitative predictions, even after invoking the ionization energy approximation. Finally, the effect of satisfying $\beta > 1$ is similar to $0 < \beta < 1$ as explained above. 

We can now extend our approach to other well-known systems, namely, CdSe/ZnS~\cite{dab,zha} and InGaAs/GaAs~\cite{sie}. Table~\ref{Table:II} lists the averaged atomic ionization energies ($E_I$) for four different core/shell structures, CdSe/ZnS, CdSe/ZnSe, CdZnSe/ZnSe and InGaAs/GaAs. In these structures, all the $E_I$ values for the cores are smaller than the shells, which in turn implies that these NCs, if graded, will feature the non-square-well potential similar to the one sketched in Fig.~\ref{fig:1}A (dashed line). If on the other hand, one switches the core(Y)/shell(Z) structure to core(Z)/shell(Y), then it is possible to obtain the non-square-well potential described in Fig.~\ref{fig:1}B (also labeled with a dashed line). 

\begin{table}[ht]
\caption{Averaged atomic ionization energies ($E_I$) for the core/shell structure of a given NC. The difference of the ionization energies between the core and shell are labeled with $|\Delta|$. See text for details.} 
\begin{tabular}{l c c c } 
\hline\hline 
\multicolumn{1}{l}{core/shell NC}   &    ~~Core        & ~~Shell            & ~~$|\Delta|$ \\  
\multicolumn{1}{l}{}                &   ~~(kJ/mol)     & ~~(kJ/mol)         & ~~(kJ/mol) \\  
\hline 

CdSe/ZnS                            &  1095 				 &  1160            &  65         \\ 
CdSe/ZnSe                           &  1095					 &  1130            &  35         \\ 
CdZnSe/ZnSe                         &  1113					 &  1130            &  17         \\
InGaAs/GaAs                         &  1787					 &  1834            &  47         \\ 

\hline  
\end{tabular}
\label{Table:II} 
\end{table}

The second issue arising from Table~\ref{Table:II} is the magnitude of $|\Delta| = |E_I^{\rm core} - E_I^{\rm shell}|$ for different NC structures. For example, one can arrange the NC structures in the order of increasing $|\Delta|$: CdZnSe/ZnSe (17 kJ/mol) $\rightarrow$ CdSe/ZnSe (35 kJ/mol) $\rightarrow$ InGaAs/GaAs (47 kJ/mol) $\rightarrow$ CdSe/ZnS (65 kJ/mol). One can use this information to further understand the required grading-depth (GD) to suppress the blinking effect in nanocrystals. The GD in this case can be defined as the average length of a Zn ion can diffuse into the core in a CdZnSe/ZnSe nanocrystal. 

\begin{figure}[hbtp!]
\begin{center}
\scalebox{0.25}{\includegraphics{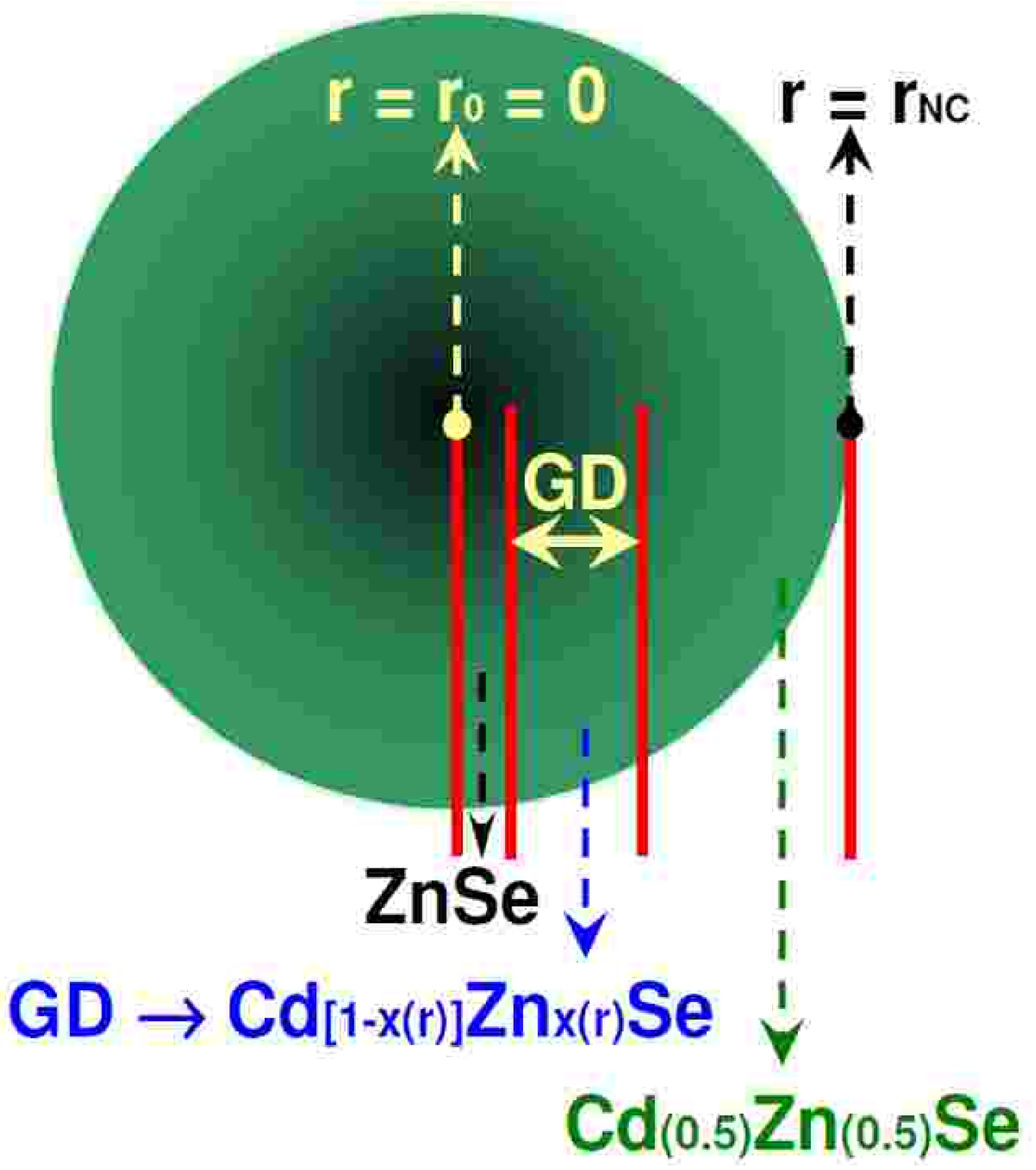}}
\caption{The definition of the grading-depth (GD) of a NC given in a two-dimensional diagrammatic form. Each region (separated with solid vertical lines) has different $x$ or Zn concentration (follow the arrows pointing downward). However, the Zn-concentration gradient [$x = x(r)$] only exist between the core, ZnSe [fixed $x$ = 1 below $r^{\rm outer}_{\rm core}$(ZnSe)] and the shell, Cd$_{0.5}$Zn$_{0.5}$Se [fixed $x$ = 0.5 above $r^{\rm inner}_{\rm shell}$(Cd$_{0.5}$Zn$_{0.5}$Se)]. Therefore, the radius of the NC satisfies the inequality, $r_0 < r^{\rm outer}_{\rm core}$(ZnSe) $< r$(Cd$_{1-x}$Zn$_{x}$Se) $< r^{\rm inner}_{\rm shell}$(Cd$_{0.5}$Zn$_{0.5}$Se) $< r_{\rm NC}$.}
\label{fig:2}
\end{center}
\end{figure}

Figure~\ref{fig:2} shows the details of the GD for the NC based on Fig.~\ref{fig:1}B. In this case, GD = $r^{\rm inner}_{\rm shell}$(Cd$_{0.5}$Zn$_{0.5}$Se) $-$ $r^{\rm outer}_{\rm core}$(ZnSe), which suggests that the gradient of Zn concentration within this GD range [sandwiched between $r^{\rm inner}_{\rm shell}$(Cd$_{0.5}$Zn$_{0.5}$Se) and $r^{\rm outer}_{\rm core}$(ZnSe)] can be large if $|\Delta|$ is made small. Otherwise, the concentration gradient has to be small. The small concentration gradient can only be achieved by having a larger GD. This latter scenario may limit the strength of the electronic confinement due to size-constraint because a large GD means a large NC size, and hence, a weak electron confinement. 

As such, it is relatively easy to suppress the blinking effect for a CdSe/ZnSe nanocrystal because the required GD is smaller, as compared to InGaAs/GaAs. For a InGaAs/GaAs nanocrystal, one needs In to diffuse far enough (smaller In concentration gradient) toward the shell to suppress the blinking effect due to large $|\Delta|$. In other words, $|\Delta|^{\rm CdSe}_{\rm ZnSe} < |\Delta|^{\rm InGaAs}_{\rm GaAs} \rightarrow 35 < 47$ kJ/mol. Smaller $|\Delta|$ also means shorter relaxation lifetimes, which are required to obtain a non-blinking NC as pointed out in Ref.~\cite{wangnat}. 

From our analysis presented above, we propose here that the origin of the blinking effect, apart from the commonly accepted trapped electrons and non-radiative processes~\cite{cpc}, could also be due to large concentration gradients quantified by the value $|\Delta|$. Indeed, large $|\Delta|$ implies longer relaxation lifetime as reported by Rajesh et al.~\cite{raj}. Note here that the gradient for an ungraded NC is infinity (large $|\Delta|$) at the interface between the core and the shell in complete agreement with our analysis thus far. Systematic grading of the NC composition splits the single-level excitation ($P \propto |\Delta|$) in an ungraded NC to a smaller multi-level excitations in a graded NC. The three-level excitations shown in Fig.~\ref{fig:1} correspond to three energy peaks, $P_1(I_1) \propto |\Delta|_1$, $P_2(I_2) \propto |\Delta|_2$ and $P_3(I_3) \propto |\Delta|_3$. Recall here that the PL energy peak intensities ($I$) are related to $x$ \textit{and therefore, to} $E_I$ for a given energy peak position ($P$), and this is how the $P$ is related to $I$. This means that if $x$ is constant, then $P(I)$ is also constant throughout the NC. Apart from that, one should also be aware that the phonons within the NC may also significantly affect the relaxation lifetimes~\cite{zibnat}. Extensive discussion on this phonon-assisted relaxation lifetime issue within the IET is given in Ref.~\cite{a4}. We emphasize here that our analysis based on the IET is complementary to the interpretations given in Ref.~\cite{wangnat}.

\subsection*{6.~~Conclusions}    

In conclusion, we have considered the variation of the ionization energy or the energy-level spacing within the nanocrystals to explain the blinking effect suppression. The ionization energy theory approximation has been employed to take this energy-level spacing variation into account. By knowing the values for the atomic ionization energy $E_I$ and the band gap difference $|\Delta|$ between the core and shell in a given nanocrystal, one can predict the possibility of achieving effective non-blinking in graded NCs and/or how to switch the randomly blinking nanocrystals into a coherently emitting NCs. Moreover, by measuring the PL spectra (both intensities and the energy peaks), one can actually understand the blinking properties of the NC by studying their non-square-well potential profiles. This information can then be used to fine-tune the composition and structure of NC materials to optimize the luminescence quality and yield. In other words, one can either completely suppress the blinking effect or, alternatively, obtain highly-controlled intermittent emission. We have also presented the possibility to use the PL spectra to understand the formation of the potential well in non-blinking NCs. In addition, one can fine-tune the single and smooth potential-well by controlling the $x(r)$ or the diffusion of Zn from the shell, ZnSe into the core, Cd$_{1-x}$Zn$_x$Se. Such graded NCs can be systematically analyzed using the PL spectra that in turn could be useful to design new non-blinking NCs and non-random-blinking NCs. The potential-well profiles for the core/shell, ZnSe/Cd$_{1-x}$Zn$_x$Se NCs and their relations with the emission intermittency can be straightforwardly verified experimentally. Finally, our approach is generic and can be applied to design a large variety of photon emitting devices, from single-photon emitters required for nanophotonics to high-intensity light emitting diodes, the ultimate light sources of the future.  

\subsection*{Acknowledgments}

A.D.A. would like to thank Kithriammal Soosay for the support. K.O. acknowledges the partial support from the Australian Research Council (ARC) and the CSIRO.

\end{document}